\documentclass{article}
\usepackage{PRIMEarxiv}

\usepackage[utf8]{inputenc} % allow utf-8 input
\usepackage[T1]{fontenc}    % use 8-bit T1 fonts
\usepackage{hyperref}       % hyperlinks
\usepackage{url}            % simple URL typesetting
\usepackage{booktabs}       % professional-quality tables
\usepackage{amsfonts}       % blackboard math symbols
\usepackage{nicefrac}       % compact symbols for 1/2, etc.
\usepackage{microtype}      % microtypography
\usepackage{lipsum}
\usepackage{fancyhdr}       % header
\usepackage{graphicx}       % graphics
\graphicspath{{media/}}     % organize your images and other figures under media/ folder

\usepackage{adjustbox}
\usepackage{tcolorbox}
\usepackage{colortbl}

\usepackage{tabularx}
\usepackage{pdflscape}

\usepackage{soul}

\title{Do We Really Even Need Data?}

\author{
  Kentaro Hoffman$^\dag$ \\
  Department of Statistics \\
  University of Washington \\
  Seattle, WA \\
  \texttt{khoffm3@uw.edu} \\
  \And
  Stephen Salerno$^\dag$ \\
  Public Health Sciences Division \\
  Fred Hutchinson Cancer Center \\
  Seattle, WA \\
  \texttt{ssalerno@fredhutch.org} \\
  \AND
  Awan Afiaz \\
  Department of Biostatistics \\
  University of Washington \\
  Public Health Sciences Division \\
  Fred Hutchinson Cancer Center \\
  Seattle, WA \\
  \texttt{aafiaz@uw.edu} \\
  \And
  Jeffrey T.~Leek$^\ddag$ \\
  Public Health Sciences Division \\
  Fred Hutchinson Cancer Center \\
  Seattle, WA \\
  \texttt{jtleek@fredhutch.org} \\
  \And
  Tyler H.~McCormick$^\ddag$\\
  Department of Statistics \\
  Department of Sociology \\
  University of Washington \\
  Seattle, WA \\
  \texttt{tylermc@uw.edu} \\
}

\begin{document}

\maketitle

\vspace{-4ex}

$^\dag$Authors contributed equally

$^\ddag$Authors contributed equally

\section*{Introduction}

\begin{figure}[!ht]
    \centering
    \includegraphics[height = 2in]{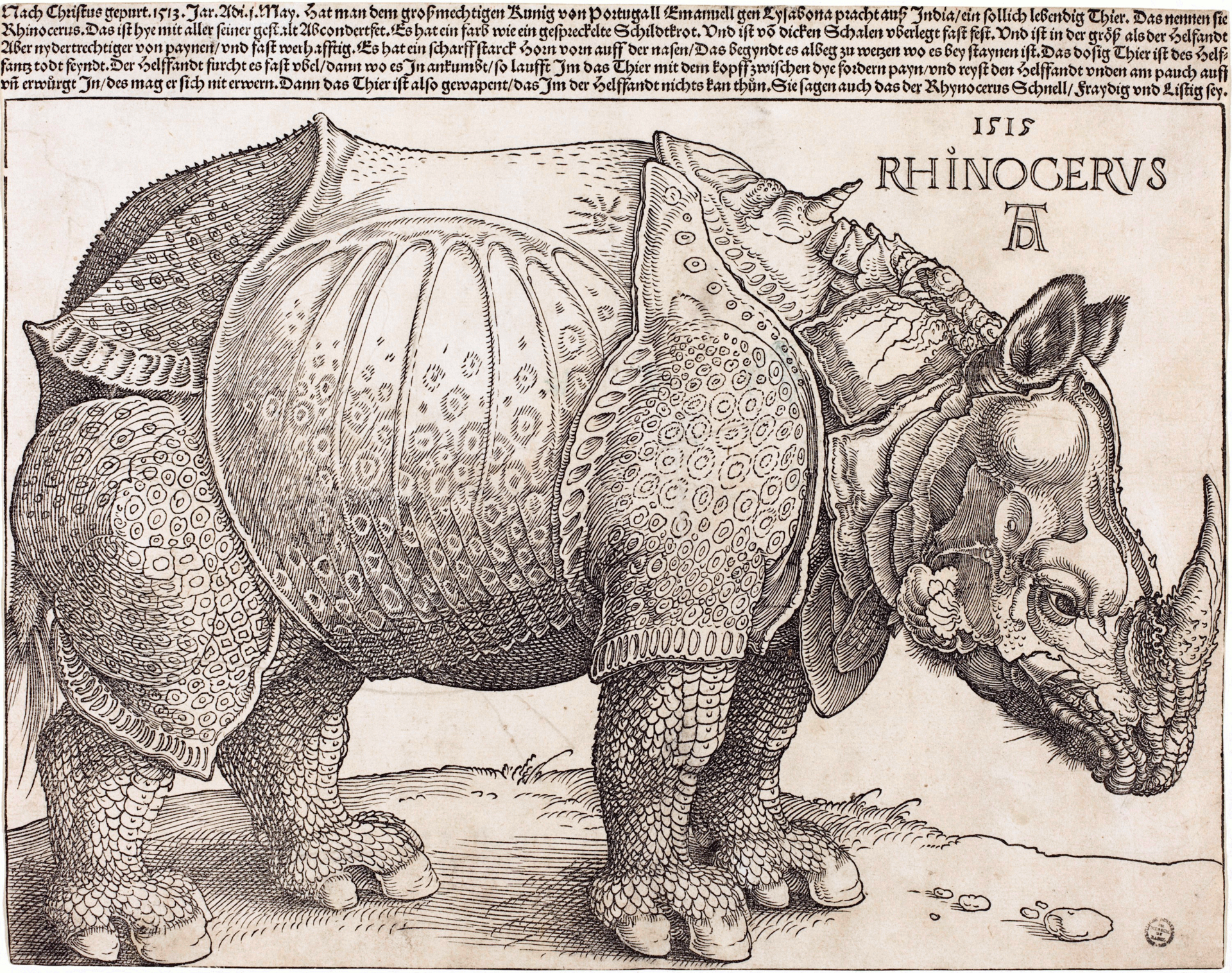}
    \includegraphics[height = 2in]{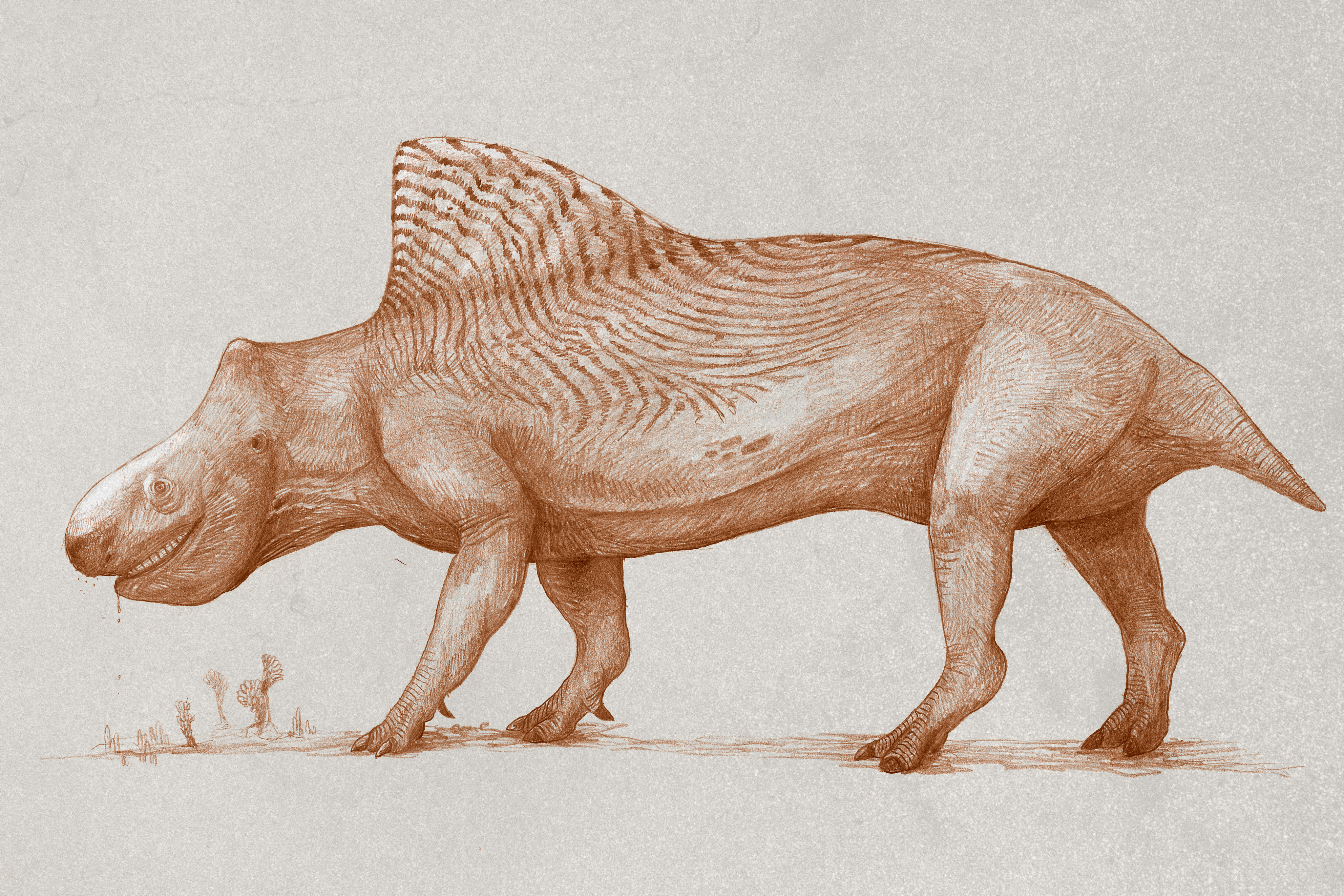}
    \caption{Artist renderings of a rhinoceros based on limited information. Left: Albrecht D{\"u}rer's {\it The Rhinoceros}, woodcutting (1515); Right: C.M.~K{\"o}semen's re-imagining of a rhinoceros based on its skeleton.}
    \label{fig:rhino}
\end{figure}

If you had never seen a rhinoceros before, how would you draw it?  Figure \ref{fig:rhino} shows two attempts. Albrecht D{\"u}rer had never seen a rhino in 1515 when he made his woodcutting (left panel). Instead, he used statistical {\it inference}. He gathered a sample of descriptions from several people who had seen a rhino to create a unified summary. While not entirely accurate, D{\"u}rer's work remained the prevailing European understanding of a rhino through the 18th century \cite{rookmaaker2005review}. 

C.M.~K{\"o}semen has seen a rhino before, but tried to imagine how we might draw one if we didn't have the luxury of first-hand reports like D{\"u}rer.  Specifically, K{\"o}semen drew his rhino the way we draw other animals no person on earth has seen, like dinosaurs~\cite{conway2012all}.  K{\"o}semen's depiction is based only on parts of the rhino that could be fossilized (right panel). It looks very different, but still matches many of the basic features of the creature we know today. K{\"o}semen's rhino, like contemporary depictions of dinosaurs, are \emph{predictions} based on potentially limited features.

The population of K{\"o}semen's rhinos is exploding across virtually all scientific settings. That is, as artificial intelligence and machine learning tools become more accessible, and scientists face new obstacles to data collection (e.g.~rising costs, declining survey response rates), researchers increasingly use predictions from pre-trained algorithms as stand-ins for hard-to-collect, and expensive samples. In genetics, phenotypes based on cheap partial genome sequences stand in for more expensive complete sequences \cite{behravan2018machine, liu2017case, gusev2019transcriptome, arumugam2011enterotypes, gamazon2015gene}. Social scientists use census data and street level photography to predict an individual's race based on the composition of their neighborhood \cite{gebru2017using}. So-called ``pragmatic'' clinical trials vastly reduce costs by predicting outcomes that may have occurred between regularly scheduled clinic visits \cite{gamerman2019pragmatic, williams2015pragmatic}. Demographers and global health researchers in low-resource settings extrapolate difficult to collect data on mortality based on small surveys and easy-to-obtain measures like light intensity \cite{evans2013estimates}. In sum, K{\"o}semen's rhino inhabits every corner of modern science. Table \ref{tab:papers_comparison} provides a list of citations to emphasize this point.

Though appealing for financial and logistical reasons, using standard tools for inference can misrepresent the association between independent variables and the outcome of interest when the true, unobserved outcome is replaced by a predicted value. The statistical framing of this problem was introduced in the context of global health and genomics \cite{wang2020methods}.  Inherent in this framework are three sources of error: (i) the relationship between predicted outcomes and their true, unobserved counterparts, (ii) the robustness of the machine learning model to resampling or uncertainty about the training data, and (iii) the ability to appropriately propagate bias {\it and} uncertainty from predictions into the ultimate inference procedure. The authors proposed estimating a correction factor using the relationship between the predicted and observed outcomes in a small, complete dataset to correct inference in subsequent datasets where the outcome is not observed. 

In this paper, we provide a description and brief overview of a rapidly growing literature focused on this general area of ``\emph{inference with predicted data (IPD)}''.  We also contrast the framework for inference with predicted data with classical work spanning several related fields, including survey sampling, missing data, and semi-supervised learning. This contrast clarifies the role of design in both classical and modern inference problems. 

\section*{The Problem}

A researcher is interested in studying an outcome, $Y$, which is difficult to measure due to practical constraints such as time or cost. They hypothesize that $Y$ is associated with $X$, a set of features which are easier to measure. Their goal is to estimate a parameter of scientific interest, $\theta$, which describes the relationship between $X$ and $Y$. To aid in this, the researchers have access to a prediction rule, $f : \mathcal{X} \to \mathcal{Y}$, mapping from the space of $\mathcal{X}$ to $\mathcal{Y}$, that was trained by an upstream group of researchers with little to no potential for information sharing. Examples of $f$ are seen in algorithms such as AlphaFold for predicting protein structures \cite{jumper2021highly}, or large language models such as ChatGPT \cite{caruccio2024can}. Given a collection of unlabeled features, where $X_{\mathcal{U}} = (X_{1, \mathcal{U}}, \ldots, X_{N, \mathcal{U}})$ is observed but $Y_{\mathcal{U}} = (Y_{1, \mathcal{U}}, \ldots, Y_{N, \mathcal{U}})$ is not, the researcher would like to use the prediction rule to impute the missing outcomes in such a way that accounts for biases in the algorithm's training procedure or biases arising from transporting the algorithm from one context to another.

IPD is a new analytic paradigm which addresses this need for drawing valid scientific inferences when an outcome is algorithmically-derived. This scenario arises primarily for two reasons: (i) it is impossible or prohibitively expensive to obtain labeled data or (ii) including data with predicted labels improves power of statistical tests. To utilize this paradigm, a measure of the relationship between the predicted and true underlying outcomes must be ascertained. To this end, current methods for IPD require the researcher to first collect a small number of labeled samples $(X_{\mathcal{L}}, Y_{\mathcal{L}}) = \{(X_{1, \mathcal{L}}, Y_{1, \mathcal{L}}) \ldots, (X_{n, \mathcal{L}}, Y_{n, \mathcal{L}})\}$. Then, by applying the prediction rule to the labeled data, the researcher can compare the predictions to the truth and ascertain the levels of bias and variability. The estimated bias level is then used to correct the parameter estimates that one would get if one used the algorithm on the unlabeled data, and the accompanying standard errors are inflated to account for the uncertainty in the predictions. 

Existing IPD methods differ based on the how the bias estimation is performed and whether the bias correction is performed on the space of outcomes or parameters \cite{wang2020methods, angelopoulos2023prediction, angelopoulos2023ppi++, miao2023assumption, egami2023using}. Wang et al.~(2020) proposed a simple IPD strategy of modeling the relationship between the predicted and observed outcomes in the labeled subset of the data. Their approach, termed \textit{post-prediction inference (PostPI)}, used this strategy as a means of estimating the true relationship between the unobserved outcomes and the features of interest. Similarly, Egami et al.~(2023) proposed a \textit{design-based semi-supervised learning approach (DSL)}, which constructs pseudo-observations based on the observed features and observed and predicted outcomes. Egami et al.~(2023) further provide mathematical guarantees as to the performance of their \textit{DSL} approach. 

\begin{figure}[!ht]
    \centering
    \begin{tcolorbox}[colback = gray!5!white, colframe = gray!75!black, title = Overview of Inference on Predicted Data]
    
    \vspace{2ex}
    
    \includegraphics[width = \textwidth]{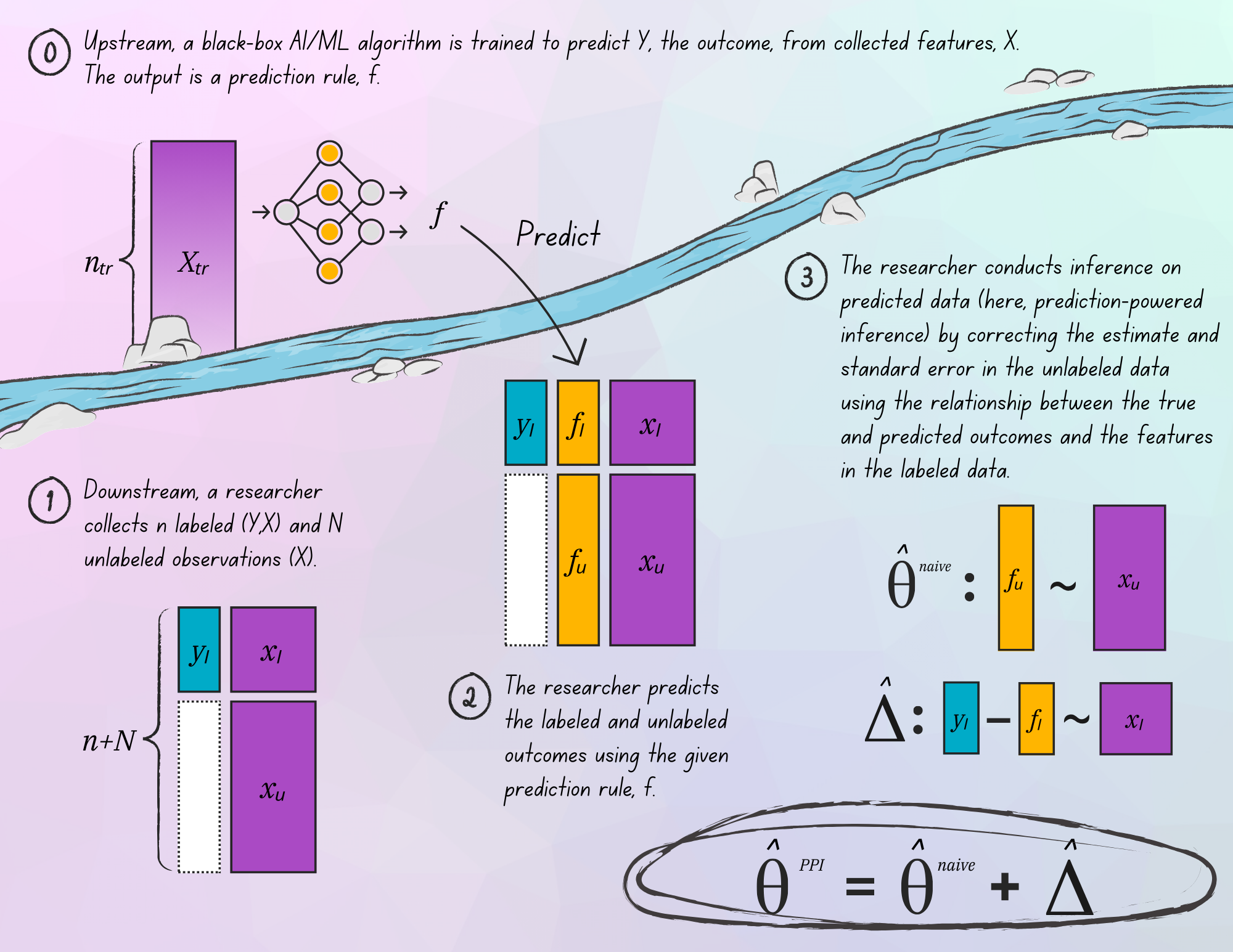}
    
    \vspace{2ex}
    
    {\bf Notes:}
    
    \begin{itemize}
        \item[-] This example overview of inference on predicted data (IPD) is based on the methodology and notation of \textit{PPI} and \textit{PPI++}, which uses a \textit{rectifier} ($\Delta$) to correct the estimation of the parameter of interest
        \item[-] Current methods for inference on predicted data differ with respect to whether the predicted outcomes are calibrated \cite{wang2020methods, egami2023using}, or the resulting parameter estimates are rectified \cite{angelopoulos2023prediction, angelopoulos2023ppi++, miao2023assumption}
        \item[-] Recent IPD methods provide theoretical guarantees so that the  inference is no worse than inference drawn on only the labeled data \cite{angelopoulos2023prediction, angelopoulos2023ppi++, egami2023using, miao2023assumption}
    \end{itemize}

    \vspace{1ex}
    
    \end{tcolorbox}
    \caption{Example general overview of \textit{inference on predicted data} (IPD) based on the \textit{Prediction-Powered Inference (PPI)} and \textit{PPI++} framework of Angelopoulos et al.~(2023a, 2023b) \cite{angelopoulos2023prediction, angelopoulos2023ppi++}}
    \label{fig:ppi_overview}
\end{figure}

Recently, the IPD was improved upon by Angelopoulos et al.~(2023a) and (2023b) with their \textit{prediction-powered inference} (PPI) and \textit{PPI++} methods and approach, respectively. These approaches instead correct the estimate of interest, such as a regression coefficient, directly, and provide theoretical guarantees that the estimates will be unbiased and that the standard errors for these estimates will be no larger than those one would obtain from using only the labeled subset of data. Miao et al.~(2023) proposed an \textit{assumption-lean and data-adaptive post-prediction inference (POP-Inf)} approach which, like \textit{PPI} and \textit{PPI++}, correct the estimated parameters themselves, but weights the use of the predicted outcomes differently for each estimated parameter. Figure \ref{fig:ppi_overview} exemplifies this concept based on the recent work of Angelopoulos et al.~(2023a) and (2023b) \cite{angelopoulos2023prediction, angelopoulos2023ppi++} whose approach has both theoretical guarantees for IPD and whose PPI notation clearly illustrates the core ideas. 

\section*{Doesn't This Sound Familiar?}

\begin{figure}[!ht]
    \centering
    \begin{tcolorbox}[colback = gray!5!white, colframe = gray!75!black, title = Alternate Methods for Analyzing Data with Missing Outcomes]
    
    \vspace{2ex}
    
    \includegraphics[width = \textwidth]{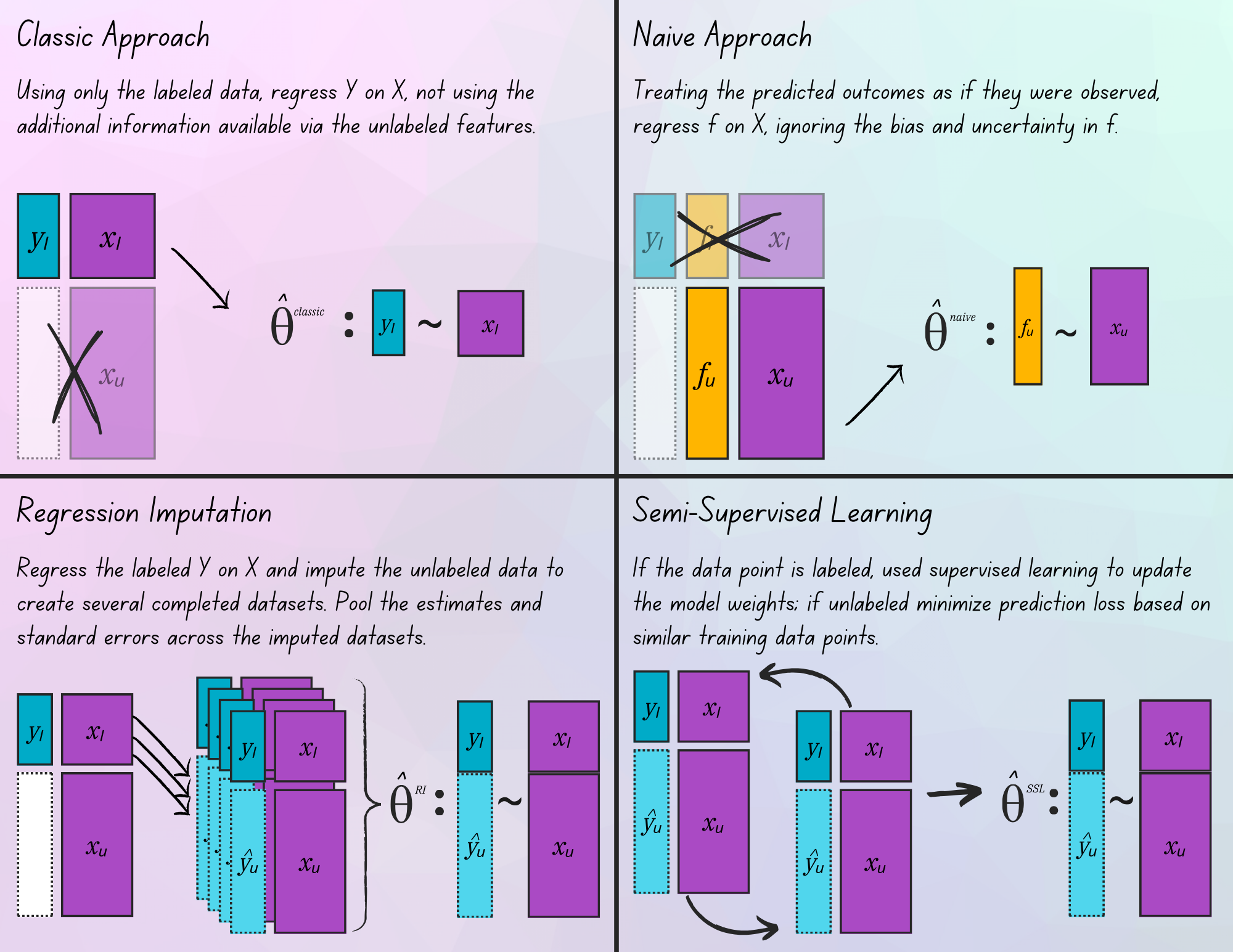}
    
    \vspace{2ex}
    
    {\bf Notes:}
    
    \begin{itemize}
        \item[-] The `classic approach' is statistically valid, but inefficient as it does not leverage additional information from $X_\mathcal{U}$ 
        \item[-] The `na\"{\i}ve approach,' which is used in many cases, may lead to bias in the estimated $\hat{\theta}$, standard errors that are too small, and anti-conservative test statistics    
        \item[-] Regression imputation assumes we can accurately estimate the missing outcomes with a model for which we choose and know the operating characteristics
        \item[-] Semi-supervised learning requires retraining the model used to predict the unobserved outcomes based on the labeled and unlabeled features, and the labeled outcomes
    \end{itemize}

    \vspace{1ex}
    
    \end{tcolorbox}
    \caption{Alternative methods for analyzing data with missing outcomes}
    \label{fig:comparison}
\end{figure}

As with other innovations, the development of methods for inference on predicted data takes inspiration and ideas from established paradigms (see Figure \ref{fig:comparison}). However the requirements of working with a pre-trained algorithm puts new restrictions on the analysis, which makes this paradigm distinct from existing approaches in two significant ways. First, is the information sharing restriction. Under IPD, it is posited that either due to propriety information (in the case of large language models), data privacy issues (medical data), feasibility of data sharing, or practicality, users of the pre-trained algorithm do not have easy access to the data the algorithm was trained on. Thus, the sample sizes required to do power calculations via two phase sampling are not available, preventing us from framing this as an auxiliary variable problem in sampling schemes. Further, the inability to ascertain the level of uncertainty of any estimated parameters makes it impossible to determine the true accuracy of missing data imputation schemes. Second, using `black box' algorithms (which include many of the exciting AI/ML developments in recent years), makes it difficult to validate the linear, parametric, or semi-parametric assumptions that underlie popular auxiliary or missing data strategies. 

\section*{Inference on Predicted Data and Current Salient Problems of Data Science}

As a paradigm that was designed to accommodate large `black box' algorithms, IPD offers a potential opportunity to address some of the most pressing aspects facing data science. First, as seen with the exponentially rising costs of fitting neural networks, the economics of training models on the most up-to-date and relevant data is becoming increasingly complex. IPD methods offer a potential way to avoid expensive refitting by learning how to calibrate results to models with reasonable baseline performance. Second, in major application domains such as healthcare, due to legal barriers concerning data privacy, some of the most relevant data contained in electronic medical records have become siloed among different organizations and corporations, hampering the uses of this information. IPD, much like the earlier development of federated learning, serves as a new way for scientists with access to different data silos to cooperate in the presence of existing data restrictions. Finally, in domains which typically rely on rigorous quantification of sources of error such as societal surveys, existing approaches of error quantification such as those based on the Total Survey Error Framework lack methods which are well suited to the evaluation of error prone, potentially `black box' models. The authors view IPD as a first pass at creating rigorous methods to create modern error quantification techniques which ensure reliability even under complex error schemes. 

\section*{Conclusions}

With the rise of artificial intelligence (AI), which trades ease of use with `black box' opacity, we are at a point where these models could massively improve our work as scientists. However, without statistically valid methods for utilizing AI/ML-derived outcomes, we run the risk of drawing our own skewed perceptions of rhinos and dinos for years to come. It is our view that inference with predicted data could serve as a powerful new toolbox to help calibrate hypothesis-driven inference across many scientific domains. More work is needed, however, to understand when and how IPD catalyzes scientific hypotheses.  If the prediction model is low quality, for example, the additional variation that comes from prediction may outweigh the benefits of including additional prediction-labeled observations.  Although in a different context (see our discussion above), evidence from over a half century of literature in the two-phase sampling literature demonstrates the challenges of incorporating predicted outcomes (see, e.g.~\cite{bose1943note,cochran1977sampling,davidov2000optimal,clark2007sampling}).  Thus, while we should celebrate the discovery of ~K{\"o}semen's rhino, we are still a long way from understanding its biological and behavioral characteristics.

\begin{landscape}
\begin{table}[!ht]
\tiny
\centering
\caption{Example papers with predicted outcomes from artificial intelligence (AI) or machine learning (ML) algorithms}
\label{tab:papers_comparison}
\vspace{2ex}
\begin{tabularx}{\linewidth}{XXXXXX}
    \toprule
    {\bf Journal (Year)} & {\bf Predicted Outcome} & {\bf Features} & {\bf Type of Algorithm} & {\bf Downstream Model}  & {\bf Citation}\\
    \midrule
    {\bf Biomedical Studies} \\
    \midrule
    Nature Scientific Reports (2018) & Case-control status for breast cancer & Single nucleotide polymorphisms & Gradient boosting (XGBoost) & Gene interaction mapping & \cite{behravan2018machine} \\
    Nature Genetics (2017) & Missing genotypes, proxy cases for disease status & Sequenced genotypes & Genotype imputation algorithm (IMPUTE2) &  Case–control associations with genetic variants (logistic regression) & \cite{liu2017case} \\
    Nature Genetics (2019) &  Gene expression & Single nucleotide polymorphisms, demographic data & Penalized regression & Transcriptome-wide association study  & \cite{gusev2019transcriptome} \\
    \midrule
    {\bf Public Health} \\
    \midrule
    Nature (2009)  & Future Google search queries related to influenza-like illnesses (ILI) & Past Google search queries related to ILI & Proprietary Google Search algorithm &  Association between forecasted search patterns and incidence of ILI & \cite{ginsberg2009detecting, lazer2014parable} \\
    Environmental Research (2021) & Fine particulate mater (PM 2.5) exposure (concentration $\times$ time) & Human movement data, PM 2.5 concentration data & Random forests & Estimated differences by race/ethnicity, income level, and urbanicity & \cite{lu2021beyond}\\
    \midrule
    {\bf Economic Studies}  \\
    \midrule
    Humanities and Social Sciences Communications (2023) & Number of shipping containers at port at a given time & Satellite images of container ports & Convolutional neural networks (U-Net) & Time series forecasting for stock returns & \cite{yu2023eye} \\
    \midrule
    {\bf Sociology Studies} \\
    \midrule
    PNAS (2017) & Motor vehicle category & Google Street View images & Convolution neural networks & Likelihood of democratic voting patterns & \cite{gebru2017using} \\
    PNAS (2017) & ``Streetscore'' measure of safety perception & Panoramic images of city streetscapes & Computer vision Algorithm & Linear regression for associations with community population and education levels & \cite{naik2017computer} \\
    \bottomrule
    {\bf Political Science} \\
    \midrule
    American Political Science Review (2019) & Text from media outlet & Text of partisan speeches & Latent Dirichlet allocation (LDA) topic model & Associations between change in media station ownership and national versus local political coverage, ideological slant of coverage, and viewership rates & \cite{martin2019local}  \\ 
    European Journal of Political Economy (2016)  & Continuous democracy index  & Political participation, civil liberty, and independence of non-government institutions & Support vector machines & Mixed effects model for association with per-capita gross domestic product & \cite{grundler2016democracy} \\
    \bottomrule
\end{tabularx}
\end{table}
\end{landscape}

%--- ACKNOWLEDGEMENTS-------------------------------------------------------------------------

\section*{Acknowledgments}

We thank C.M.~K{\"o}semen for the gracious use of the paleoart image. Additionally, we thank Professor Michael Jordan for his valuable suggestions, which have greatly improved the quality of this work. We would also like to express our gratitude to the co-authors of the \textit{DSL}, Naoki Egami, Musashi Jacobs-Harukawa, Brandon M Stewart, and Hanying Wei, \textit{PPI} and \textit{PPI++}, Anastasios N. Angelopoulos, Stephen Bates, Clara Fannjiang, John C. Duchi, Michael I. Jordan, and Tijana Zrnic, and \textit{POP-Inf}, Jiacheng Miao, Xinran Miao, Yixuan Wu, Jiwei Zhao, and Qiongshi Lu, methods for their contributions and collaborative efforts in the development of this field.

%--- BIBLIOGRAPHY ----------------------------------------------------------------------------

\bibliographystyle{unsrt}  
\bibliography{references}  

\begin{thebibliography}{10}

\bibitem{rookmaaker2005review}
Leendert~C Rookmaaker.
\newblock Review of the european perception of the african rhinoceros.
\newblock {\em Journal of Zoology}, 265(4):365--376, 2005.

\bibitem{conway2012all}
John Conway, C.M. Kosemen, Darren Naish, and Scott Hartman.
\newblock {\em All yesterdays: Unique and speculative views of dinosaurs and others prehistoric animals}.
\newblock Irregular books, 2012.

\bibitem{behravan2018machine}
Hamid Behravan, Jaana~M Hartikainen, Maria Tengstr{\"o}m, Katri Pylk{\"a}s, Robert Winqvist, Veli-Matti Kosma, and Arto Mannermaa.
\newblock Machine learning identifies interacting genetic variants contributing to breast cancer risk: A case study in finnish cases and controls.
\newblock {\em Scientific Reports}, 8(1):13149, 2018.

\bibitem{liu2017case}
Jimmy~Z Liu, Yaniv Erlich, and Joseph~K Pickrell.
\newblock Case--control association mapping by proxy using family history of disease.
\newblock {\em Nature Genetics}, 49(3):325--331, 2017.

\bibitem{gusev2019transcriptome}
Alexander Gusev, Kate Lawrenson, Xianzhi Lin, Paulo~C Lyra~Jr, Siddhartha Kar, Kevin~C Vavra, Felipe Segato, Marcos~AS Fonseca, Janet~M Lee, Tanya Pejovic, et~al.
\newblock A transcriptome-wide association study of high-grade serous epithelial ovarian cancer identifies new susceptibility genes and splice variants.
\newblock {\em Nature Genetics}, 51(5):815--823, 2019.

\bibitem{arumugam2011enterotypes}
Manimozhiyan Arumugam, Jeroen Raes, Eric Pelletier, Denis Le~Paslier, Takuji Yamada, Daniel~R Mende, Gabriel~R Fernandes, Julien Tap, Thomas Bruls, Jean-Michel Batto, et~al.
\newblock Enterotypes of the human gut microbiome.
\newblock {\em Nature}, 473(7346):174--180, 2011.

\bibitem{gamazon2015gene}
Eric~R Gamazon, Heather~E Wheeler, Kaanan~P Shah, Sahar~V Mozaffari, Keston Aquino-Michaels, Robert~J Carroll, Anne~E Eyler, Joshua~C Denny, GTEx Consortium, Dan~L Nicolae, et~al.
\newblock A gene-based association method for mapping traits using reference transcriptome data.
\newblock {\em Nature genetics}, 47(9):1091--1098, 2015.

\bibitem{gebru2017using}
Timnit Gebru, Jonathan Krause, Yilun Wang, Duyun Chen, Jia Deng, Erez~Lieberman Aiden, and Li~Fei-Fei.
\newblock Using deep learning and google street view to estimate the demographic makeup of neighborhoods across the united states.
\newblock {\em Proceedings of the National Academy of Sciences}, 114(50):13108--13113, 2017.

\bibitem{gamerman2019pragmatic}
Victoria Gamerman, Tianxi Cai, and Amelie Els{\"a}{\ss}er.
\newblock Pragmatic randomized clinical trials: best practices and statistical guidance.
\newblock {\em Health Services and Outcomes Research Methodology}, 19:23--35, 2019.

\bibitem{williams2015pragmatic}
Hywel~C Williams, Esther Burden-Teh, and Andrew~J Nunn.
\newblock What is a pragmatic clinical trial.
\newblock {\em J Invest Dermatol}, 135(6):1--3, 2015.

\bibitem{evans2013estimates}
Jessica Evans, Aaron van Donkelaar, Randall~V Martin, Richard Burnett, Daniel~G Rainham, Nicholas~J Birkett, and Daniel Krewski.
\newblock Estimates of global mortality attributable to particulate air pollution using satellite imagery.
\newblock {\em Environmental Research}, 120:33--42, 2013.

\bibitem{wang2020methods}
Siruo Wang, Tyler~H McCormick, and Jeffrey~T Leek.
\newblock Methods for correcting inference based on outcomes predicted by machine learning.
\newblock {\em Proceedings of the National Academy of Sciences}, 117(48):30266--30275, 2020.

\bibitem{jumper2021highly}
John Jumper, Richard Evans, Alexander Pritzel, Tim Green, Michael Figurnov, Olaf Ronneberger, Kathryn Tunyasuvunakool, Russ Bates, Augustin {\v{Z}}{\'\i}dek, Anna Potapenko, et~al.
\newblock Highly accurate protein structure prediction with alphafold.
\newblock {\em Nature}, 596(7873):583--589, 2021.

\bibitem{caruccio2024can}
Loredana Caruccio, Stefano Cirillo, Giuseppe Polese, Giandomenico Solimando, Shanmugam Sundaramurthy, and Genoveffa Tortora.
\newblock Can chatgpt provide intelligent diagnoses? a comparative study between predictive models and chatgpt to define a new medical diagnostic bot.
\newblock {\em Expert Systems with Applications}, 235:121186, 2024.

\bibitem{angelopoulos2023prediction}
Anastasios~N. Angelopoulos, Stephen Bates, Clara Fannjiang, Michael~I. Jordan, and Tijana Zrnic.
\newblock Prediction-powered inference.
\newblock {\em Science}, 382(6671):669--674, 2023.

\bibitem{angelopoulos2023ppi++}
Anastasios~N Angelopoulos, John~C Duchi, and Tijana Zrnic.
\newblock Ppi++: Efficient prediction-powered inference.
\newblock {\em arXiv preprint arXiv:2311.01453}, 2023.

\bibitem{miao2023assumption}
Jiacheng Miao, Xinran Miao, Yixuan Wu, Jiwei Zhao, and Qiongshi Lu.
\newblock Assumption-lean and data-adaptive post-prediction inference.
\newblock {\em arXiv preprint arXiv:2311.14220}, 2023.

\bibitem{egami2023using}
Naoki Egami, Musashi Jacobs-Harukawa, Brandon~M Stewart, and Hanying Wei.
\newblock Using large language model annotations for valid downstream statistical inference in social science: Design-based semi-supervised learning.
\newblock {\em arXiv preprint arXiv:2306.04746}, 2023.

\bibitem{bose1943note}
Chameli Bose.
\newblock Note on the sampling error in the method of double sampling.
\newblock {\em Sankhya}, 6(3):329–30, 1943.

\bibitem{cochran1977sampling}
William~Gemmell Cochran.
\newblock {\em Sampling techniques}.
\newblock john wiley \& sons, 1977.

\bibitem{davidov2000optimal}
Ori Davidov and Yoel Haitovsky.
\newblock Optimal design for double sampling with continuous outcomes.
\newblock {\em Journal of Statistical planning and Inference}, 86(1):253--263, 2000.

\bibitem{clark2007sampling}
Robert~G Clark and David~G Steel.
\newblock Sampling within households in household surveys.
\newblock {\em Journal of the Royal Statistical Society Series A: Statistics in Society}, 170(1):63--82, 2007.

\bibitem{ginsberg2009detecting}
Jeremy Ginsberg, Matthew~H Mohebbi, Rajan~S Patel, Lynnette Brammer, Mark~S Smolinski, and Larry Brilliant.
\newblock Detecting influenza epidemics using search engine query data.
\newblock {\em Nature}, 457(7232):1012--1014, 2009.

\bibitem{lazer2014parable}
David Lazer, Ryan Kennedy, Gary King, and Alessandro Vespignani.
\newblock The parable of google flu: traps in big data analysis.
\newblock {\em science}, 343(6176):1203--1205, 2014.

\bibitem{lu2021beyond}
Yougeng Lu.
\newblock Beyond air pollution at home: Assessment of personal exposure to pm2. 5 using activity-based travel demand model and low-cost air sensor network data.
\newblock {\em Environmental Research}, 201:111549, 2021.

\bibitem{yu2023eye}
Honghai Yu, Xianfeng Hao, Liangyu Wu, Yuqi Zhao, and Yudong Wang.
\newblock Eye in outer space: satellite imageries of container ports can predict world stock returns.
\newblock {\em Humanities and Social Sciences Communications}, 10(1):1--16, 2023.

\bibitem{naik2017computer}
Nikhil Naik, Scott~Duke Kominers, Ramesh Raskar, Edward~L Glaeser, and C{\'e}sar~A Hidalgo.
\newblock Computer vision uncovers predictors of physical urban change.
\newblock {\em Proceedings of the National Academy of Sciences}, 114(29):7571--7576, 2017.

\bibitem{martin2019local}
Gregory~J Martin and Joshua McCrain.
\newblock Local news and national politics.
\newblock {\em American Political Science Review}, 113(2):372--384, 2019.

\bibitem{grundler2016democracy}
Klaus Gr{\"u}ndler and Tommy Krieger.
\newblock Democracy and growth: Evidence from a machine learning indicator.
\newblock {\em European journal of political economy}, 45:85--107, 2016.

\end{thebibliography}

%=== END =====================================================================================

\end{document}